\documentclass[11pt]{article}

\usepackage[dvips]{graphicx}

\begin{document}

\begin{titlepage}

\noindent
\hfill LIGO-P030056

\begin{center}

\vfill
{\Large\bf The Response of Test Masses to Gravitational Waves 
in the Coordinates of a Local Observer}
\vspace{1cm}

{Malik~Rakhmanov}

\vspace{0.5cm}
{\it Department of Physics, University of Florida, Gainesville, FL 32611}
\footnote{Email: malik@phys.ufl.edu}

\end{center}

\vfill
\begin{abstract}
The response of laser interferometers to gravitational waves has 
been calculated in a number of different ways, particularly in the
transverse-traceless and the local Lorentz gauges. At first sight, it
would appear that these calculations lead to different results when
the separation between the test masses becomes comparable to the
wavelength of the gravitational wave. In this paper this discrepancy
is resolved. We describe the response of free test masses to plane
gravitational waves in the coordinate frame of a local observer and
show that it acquires contributions from three different effects: 
the displacement of the test masses, the apparent change in the photon
velocity, and the variation in the clock speed of the local observer, 
all of which are induced by the gravitational wave. Only when taken 
together do these three effects represent a quantity which is 
translationally invariant. This translationally-invariant quantity 
is identical to the response function calculated in the 
transverse-traceless gauge. We thus resolve the well-known discrepancy 
between the two coordinates systems, and show that the results found 
in the coordinate frame of a local observer are valid for large 
separation between the masses.
\end{abstract}

\vfill
\end{titlepage}

\section{Introduction}

Searches for gravitational waves are now conducted with laser
interferometers in which the test masses for sensing gravitational
waves are separated by distances of several kilometers
\cite{Barish:1999, Bradaschia:1990}. Variations in the proper distance
between these test masses, which may be caused by gravitational
waves, are measured with light. The response of the laser
interferometers to gravitational waves is usually calculated in the
transverse-traceless (TT) gauge \cite{Misner:1973}. The main
assumption for all such calculations is that the test masses of the
laser interferometers are inertial, i.e. accelerations of the test
masses in the direction of the probe laser beam are negligible. A
substantial engineering effort has been made to meet this
requirement. Placed in ultra-high vacuum and isolated from the ground
by multi-layer stacks and  actively-controlled suspensions, these test
masses become practically inertial at frequencies far above the
suspension resonances.

A significant change in the attitude toward the laser interferometers
took place with the introduction of optical springs in the last few
years \cite{Buonanno:2001, Khalili:2001, Rakhmanov:PhD}. The optical
spring is produced by the pressure of light on the test masses, which
in the advanced interferometer configurations can lead to
amplification of the gravitational wave signal. For this
amplification the resonance frequency of the optical spring must be
matched with the frequency of the expected gravitational waves. In
this case, the main assumption of the TT-gauge -- the requirement of
test mass inertiality -- can no longer be made. This problem can be
overcome if one uses the coordinates of a local observer for which no
requirement of test mass inertiality is needed. Although this
coordinate system has been frequently used to describe the response
of resonant bar detectors \cite{Weber:1961}, its application to laser
interferometers thus far has been only  occasional.

The coordinates of a local observer, also known as the local Lorentz
gauge, have a long history. Comparison of the TT coordinates and the
coordinates of a local observer is given in \cite{Misner:1973} with
the curious observation that they yield different answers for
geodesic deviation  when separation between the geodesics becomes
comparable to the wavelength of the gravitational wave (Exercise 37.6
in \cite{Misner:1973}). As a result, the coordinates of a local
observer were considered not suitable for large separation between
the geodesics. Nonetheless, the studies of the effects of
gravitational waves in the coordinates of a local observer continued
and over the years led to a number of interesting results. Several
insightful papers have been written about the role of the coordinate
system in the detection of gravitational waves \cite{Grishchuk:1977,
Pegoraro:1978, Fortini:1982, Flores:1986}. Some of the calculations in
these papers rely on the Fermi normal expansion as a means to build
the coordinate frame of a local observer. Explicit transformations
from the TT-coordinates to the coordinates of a local observer have
been constructed and analyzed \cite{Grishchuk:1980, Fortini:1990,
Callegari:1987}. The response of the interferometric gravitational
wave detector calculated in the TT-coordinates was transformed into
the coordinates of the local observer \cite{Callegari:1987,
Fortini:1991}.

Despite of all these efforts, the coordinates of a local observer have
remained an obscure gauge even to this day. One of the reasons for the
lack of understanding is the avoidance of the local Lorentz gauge
which is largely influenced by the  disagreement between this gauge
and the TT-gauge. Another drawback associated with the coordinates of
a local observer is the lack of consistent mathematical formalism.  To
derive any nontrivial result in these coordinates, one usually starts
with TT-gauge and then obtains the answer by complicated coordinate
transformations. In this paper we show that it is possible to
calculate  the effects of the gravitational wave directly in the
coordinates of a local observer. There is no need to start with the
TT-gauge and no need to use the transformation rules to go from one
coordinate frame to the other. In particular, we describe the response
of test masses to gravitational waves when their separation is
comparable to or greater than the wavelength of the gravitational
wave. In this approach, several effects must be combined to obtain a
consistent test mass response. In the end, the discrepancy between the
two gauges is resolved.

The presentation in this paper is such that only a few concepts from
differential geometry are used. Often, abstract mathematical
derivations are replaced with those based on simple physical
arguments, and many formulas are deliberately presented in the
Newtonian form after they have been derived in general relativity.
The motivation for this approach is two-fold. On one hand, it allows
us to focus on physics of the problem and set aside  mathematical
details which can be overwhelming.  On the other hand, such an
approach allows us to assume the standpoint of a ``Newtonian
physicist'' \cite{Misner:1973}, conducting experiments in a 
laboratory environment and describing the outcomes of these
experiments in the familiar Newtonian terms, even though they
represent the effects in general relativity.

\section{The Coordinates of Transverse Traceless Gauge}

We begin with a brief overview of the TT-gauge. This digression will 
allow us to introduce the test mass response function which will be
needed later for comparison. Subsequent calculations, however, do not
rely on the TT-gauge in any way.

In the TT-gauge the metric which describes a plane polarized 
gravitational wave propagating in flat space-time is given by
\begin{equation}\label{metricTT}
   g_{\mu \nu} = \left(
   \begin{array}{cccc}
   -1  &             &             &    \\
       &  1\! +\! h  &             &    \\
       &             &  1\! -\! h  &    \\
       &             &             &  1 
   \end{array}\right) ,
\end{equation}
where $h = h(t + z/c)$ represents the amplitude of the ``+'' 
polarization \cite{Misner:1973}. For all anticipated astrophysical 
sources, the amplitude of gravitational waves upon their arrival to 
Earth is expected to be extremely small, typically $|h| \sim 10^{-21}$ 
or less, which justifies the use of the perturbation method in the
following calculations.

\begin{figure}[t]
   \centering\includegraphics[width=0.7\textwidth]{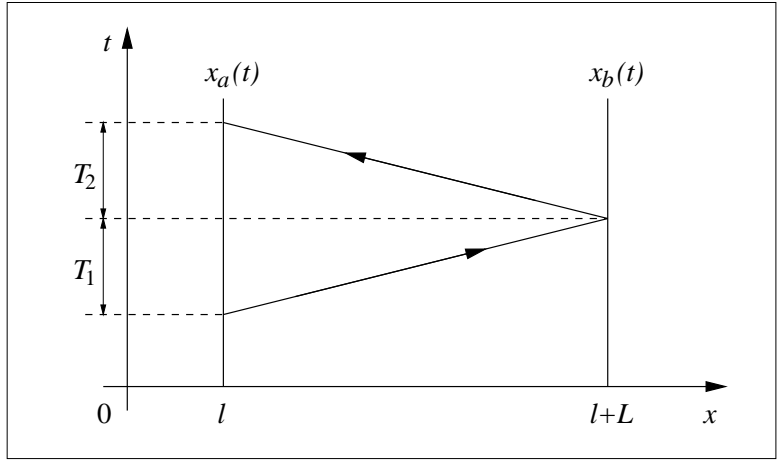}
   \caption{Bouncing photon in the coordinates of the TT-gauge.} 
   \label{rtripTT}
\end{figure}

A special property of the TT-coordinates is that an inertial test
mass, which is initially at rest in these coordinates, remains at
rest throughout the entire passage of the gravitational wave
\cite{Misner:1973, Schutz:1985}.  Here, the use of words ``at rest''
requires clarification: they only mean that the coordinates of the
test mass do not change in the presence of the gravitational wave. The
proper distance between any two test masses changes even though their
coordinates remain the same. A convenient way to analyze variations
in the proper distance is by means of ``bouncing photons''
\cite{Synge:1960}. For example, a photon can be launched from one
test mass to be bounced back by the other, as shown in 
Fig.~\ref{rtripTT}. For simplicity we assume that the test masses are
located along the $x$-axis of the coordinate system. In this case, the
interval takes the form:
\begin{equation}
   ds^2 = - c^2 \, d t^2 + [1 + h(t)] \, d x^2.
\end{equation}
The condition for a null trajectory ($ds = 0$) gives us the coordinate
velocity of the photon:
\begin{equation}
   v^2 \equiv \left( \frac{dx}{dt} \right)^2 = \frac{c^2}{1 + h(t)},
\end{equation}
which is a convenient quantity for calculations of the photon
propagation times between the test masses. As we know, the
coordinates of the test masses, $x_a = l$ and $x_b = l + L$, do not 
change under the influence of gravitational wave. Therefore, the 
duration of the forward trip can be found as
\begin{equation}
   T_1(t) = \int\limits_{l}^{l+L} \frac{dx}{v(t')},
\end{equation}
where $t' = t - (l + L - x)/c$. To first order in $h$, this integral
can be approximated as
\begin{equation}
   T_1(t) = T + \frac{1}{2c} \int\limits_{l}^{l+L} h(t') \, dx ,
\end{equation}
where $T = L/c$ is the light transit time in the absence of
gravitational waves. Similarly, the duration of the return trip would be
\begin{equation}
   T_2(t) = T + \frac{1}{2c} \int\limits_{l+L}^{l} h(t') \, (-dx) ,
\end{equation}
though now the retardation time is given by $t' = t - (x - l)/c$.

The round trip time can then be found by adding $T_2(t)$ and
$T_1[t-T_2(t)]$. The latter can be approximated by $T_1(t-T)$ because 
the difference between the exact and the approximate values is second 
order in $h$. Therefore, to first order in $h$, the duration of the 
round trip can be defined as
\begin{equation}
   T_{\mathrm{r.t.}}(t) = T_1(t - T) + T_2(t).
\end{equation}
Deviations of this round-trip time from its unperturbed value, 
$2T$, are then given by
\begin{equation}\label{delayTT}
   \delta T(t) = \frac{1}{2c} \int\limits_l^{l + L} \left[ \,
      h \left(t - 2T + \frac{x-l}{c} \right) + 
      h \left(t - \frac{x-l}{c} \right) \, \right] dx .
\end{equation}
Even though $l$ explicitly enters this equation, $\delta T$ does not 
depend on $l$. This observation implies that the choice of the origin 
for this coordinate system does not affect $\delta T$; in other
words, the result is translationally invariant.

The deviation in the round-trip time, Eq.(\ref{delayTT}), can also be 
written in the Laplace or Fourier domain. Laplace transformations are
commonly used to analyze linear responses of interferometric
gravitational-wave detectors \cite{Mizuno:1997}, and sometimes are
easier to interpret than their time-domain equivalents. Define the 
Laplace transform of an arbitrary function of time $h(t)$ by
\begin{equation}
   \tilde{h}(s) = \int \limits_0^{\infty} e^{-st} \, h(t) \; dt .
\end{equation}
Then the Laplace-domain version of Eq.(\ref{delayTT}) can be written as
\begin{equation}\label{dTinTTh}
   \frac{\delta \tilde{T}(s)}{T} = C(s) \, \tilde{h}(s),
\end{equation}
where $C(s)$ represents the response of test masses to gravitational 
waves:
\begin{equation}\label{respTT}
   C(s) = \frac{1 - e^{-2sT}}{2sT} .
\end{equation}
A number of derivations of this result, some quite different from
ours, can be found in literature, for example in
\cite{Estabrook:1975, Estabrook:1985, Gursel:1984, Vinet:1986} 
and more recently in \cite{Saulson:1994, Mizuno:1997}.

There are several reasons why the above picture is not satisfactory
from a physical point of view, even though it is mathematically
sound. The main problem with the coordinates of the TT-gauge is that
they can hardly be realized in the experiment. In fact, they cannot
be implemented in the laboratory environment on Earth because the
coordinate grid of the TT-frame must be changing in unison with the
passing gravitational wave, the effect commonly known as ``breathing
of the frame.'' (They may, however, be realized in space with a
network of freely-falling satellites.) Consequently, the application
of the above calculations to ground-based  gravitational-wave
detectors becomes problematic. For physicists working with these
detectors, it is sometimes not clear how the results derived in the
TT-coordinates can be used in experiments when these coordinates are
not available in practice.

Another problem, which is closely related to the previous one, comes
from the assumption of test mass inertiality. Namely, the above
derivation of the photon round-trip time was based on the premise that
the coordinates of the test masses do not change under the influence
of the gravitational wave, an assumption which is true only when the
test masses are inertial. The test masses in laser gravitational-wave
detectors constantly undergo accelerations in response to various
forces acting on them and thus are never truly inertial. One can
argue that these accelerations typically occur at frequencies of the
suspension resonances which are well below the frequencies of the
anticipated gravitational waves. However, in advanced interferometer
configurations the accelerations of test masses will also be caused by
the radiation-pressure variations which are intended to occur at the
frequency of anticipated gravitational waves. Therefore, the
assumption of test mass inertiality, which is most effective in the
TT-gauge, becomes too restrictive for more realistic calculations. 
These problems do not occur if one uses the coordinates of a local 
observer.

\section{The Coordinates of Local Observer}

An observer in a laboratory environment on Earth typically uses the
coordinate system in which the space-time is locally flat
\cite{Synge:1960}, and the distance between any two points is given
simply by the difference in their coordinates in the usual sense of
Newtonian physics \cite{Misner:1973}. In this reference frame,
gravitational waves manifest themselves through the tidal forces
which they exert on the masses \cite{Weber:1961}. To describe the
tidal forces we consider a test mass which is free to move in the
horizontal plane ($z = 0$). For simplicity, we assume that this plane
coincides with the wavefront of the gravitational wave, and that the
$x$ and $y$ directions of the coordinate  system match the
polarization of the gravitational wave. Then the tidal acceleration of
the test mass caused by the gravitational wave \cite{Misner:1973} is
given by
\begin{eqnarray}
   \ddot{x} & = & + \frac{1}{2} \; \ddot{h} \; x ,
                  \label{gwAccelX} \\
   \ddot{y} & = & - \frac{1}{2} \; \ddot{h} \; y .
                  \label{gwAccelY} 
\end{eqnarray}
Equivalently \cite{Grishchuk:1977, Grishchuk:1980, Blandford:2003}, 
one can say that there is a gravitational potential:
\begin{equation}\label{potential}
   \Phi({\mathbf{r}}, t) = - \frac{1}{4} \, \ddot{h}(t) (x^2 - y^2) ,
\end{equation}
which generates the tidal forces, and that the motion of the test mass
is governed by the Newton equation:
\begin{equation}\label{NewtonLaw}
   \ddot{\mathbf{r}} = - \nabla \Phi .
\end{equation}
The potential is not static, and therefore the energy of the test mass
is not conserved. This is the Newtonian version of the theorem from
general relativity which states that gravitational waves must supply 
energy to test masses to become detectable in experiments
\cite{Bondi:1959}.

In the post-Newtonian approach to general relativity (see also
Appendix \ref{App:TT2FN}), the gravitational potential is related to
the time component of the metric:
\begin{equation}\label{g00postN}
   g_{00} = - 1 - \frac{2}{c^2} \; \Phi .
\end{equation}
In what follows we will frequently use a perturbation expansion,
keeping only terms which are first order in $h$, and therefore rely on 
the assumption that $|\Phi|/c^2 \ll 1$. To satisfy this condition, we 
require that the spatial coordinates $x$ and $y$ do not extend
indefinitely. This limitation, however, will not restrict us in any way. 
Indeed, for gravitational waves with the largest expected amplitudes
($|h| \sim 10^{-21}$) and the highest detectable frequencies
($\sim 10$ kHz), the restriction on the spatial coordinates implies
that $|x|, |y| \ll 10^{14}$~m, which is always satisfied in the
laboratory environment on Earth.

The solution to Eqs.(\ref{gwAccelX})--(\ref{gwAccelY}) is usually
found using the perturbation method \cite{Misner:1973}. To first order
in $h$, the displacements of the test mass caused by the gravitational 
waves are given by
\begin{eqnarray}
   \delta x(t) & = & + \frac{1}{2} \; x_0 \; h(t) ,
                    \label{shiftXgw} \\
   \delta y(t) & = & - \frac{1}{2} \; y_0 \; h(t) ,
                    \label{shiftYgw} 
\end{eqnarray}
where $x_0$ and $y_0$ are the initial (unperturbed) coordinates of the 
test mass. This notion is regarded as the major difference between the 
coordinates of a local observer and the coordinates of the TT-gauge,
in which the test masses were not moving under the influence of the
gravitational wave.

\section{Requirement of Translational Invariance}

An interesting feature of the local Lorentz gauge is the coordinate 
dependence of the tidal forces -- they can be changed by a mere shift 
of the origin of the coordinate system: 
\begin{equation}\label{shiftsXY}
   x \rightarrow x + X, \qquad {\mathrm{and}} \qquad 
   y \rightarrow y + Y.
\end{equation}
The same applies to the test mass displacements, Eqs.(\ref{shiftXgw}) 
and (\ref{shiftYgw}). This is the earliest indication that the 
coordinates of a local observer are not as simple as they may seem.
However, at this point, the coordinate dependence seems to be quite 
harmless, and we can entertain the notion that it can be removed 
simply by considering the relative motion of test masses.

As before, we probe the geometry of space-time with a bouncing photon.
Consider two test masses with coordinates $x_a$ and $x_b$ and assume
that the photon is launched from one test mass and is bounced by the
other. Let the unperturbed values for the test mass coordinates be
\begin{equation}\label{XaXb2}
   x_a = l , \qquad {\mathrm{and}} \qquad 
   x_b = l + L ,
\end{equation}
and the unperturbed propagation time between the masses be
\begin{equation}
   T = \frac{L}{c} .
\end{equation}
From Eq.(\ref{shiftXgw}) we find that the displacements of the test 
masses under the influence of the gravitational wave are
\begin{eqnarray}
   \delta x_a(t) & = & \frac{1}{2} \; l \; h(t) ,\\
   \delta x_b(t) & = & \frac{1}{2} \; (l + L) \; h(t) .
\end{eqnarray}
Then the relative displacement, commonly defined as
\begin{eqnarray}
   \delta L(t) & = & \delta x_b(t) - \delta x_a(t) 
                     \label{dLinst} \nonumber \\
               & = & \frac{1}{2} \; L \, h(t) ,
                     \label{unphysDL} 
\end{eqnarray}
would obviously be independent of $l$ and therefore independent of
the choice of the origin for these coordinates, as we anticipated. 
Equation (\ref{unphysDL}), often written as
\begin{equation}\label{strain_h(t)}
   \frac{\delta L(t)}{L} = \frac{1}{2} \; h(t) ,
\end{equation}
is widely used to describe the strain induced by gravitational
waves on bar detectors. However, its application to laser
interferometers immediately runs into a problem. Namely, the change 
in the round-trip time calculated from Eq.(\ref{strain_h(t)}) would be
\begin{equation}\label{naiveDT}
   \frac{\delta \tilde{T}(s)}{T} = \tilde{h}(s) ,
\end{equation}
which is different from the one obtained in the TT-gauge,
Eq.(\ref{dTinTTh}). This is the precise origin of the well-known
discrepancy between the two coordinate systems. One of the earliest
accounts of this discrepancy appears in Exercise 37.6 of
\cite{Misner:1973}, which also suggests that the correct answer for
the photon propagation time must be obtained in the coordinates of
the TT-gauge. It is sometimes assumed that the discrepancy occurred
because of the application of Eqs.(\ref{gwAccelX})--(\ref{gwAccelY})
beyond their limits of validity. The actual cause of the discrepancy
lies in the neglect of the effects of gravitational redshift, as will
be shown below.

Historically, the discrepancy was not viewed as a serious problem 
when the searches for gravitational waves were conducted with bar
detectors. The relatively small size of a bar detector (a few meters)
implies small separation for its constituent parts, in which case the
difference between the two coordinate systems becomes negligible.
Indeed, for gravitational waves with wavelength much greater than 
the separation between the test masses, $|sT| \ll 1$ and therefore 
$C(s) \approx 1$, which makes Eq.(\ref{dTinTTh}) equivalent to
Eq.(\ref{naiveDT}). The situation became rather different with the
arrival of long-baseline laser interferometers. In these detectors the
test masses for sensing gravitational waves are separated by distances
of several kilometers, and the long-wavelength approximation, 
$|sT| \ll 1$, becomes hard to justify. Furthermore, recent studies
\cite{Markowicz:2003} have shown that these interferometers are
capable of detecting gravitational waves with wavelengths comparable
to their arm-length, $|sT|\sim1$, thus operating entirely outside the
long-wavelength regime.

\section{Requirement of Causality}

\begin{figure}[t]
   \centering\includegraphics[width=0.7\textwidth]{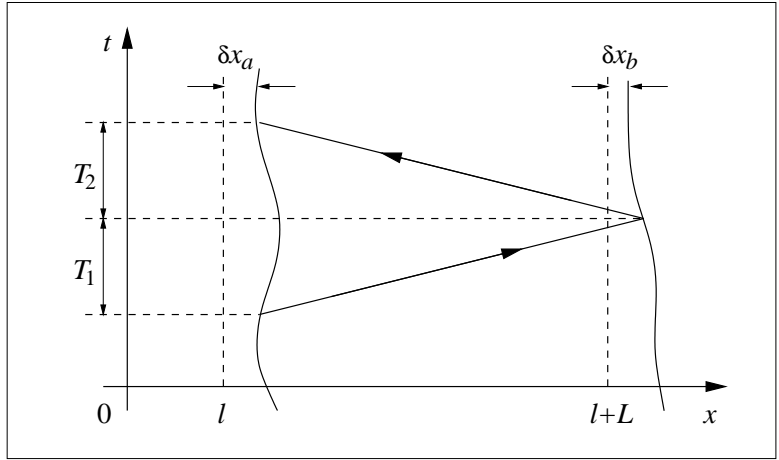}
   \caption{Bouncing photon in the coordinates of a local observer.} 
   \label{rtripLL}
\end{figure}

For large separation between the test masses, the definition for 
relative displacement, Eq.(\ref{dLinst}), becomes unphysical. In this 
definition the two test masses are taken at the same time and
therefore cannot be in causal connection. The definitions for 
the relative test-mass displacement which are appropriate for the 
bouncing photon can be written as
\begin{eqnarray}
   \delta L_1(t) & = & \delta x_b(t) - \delta x_a(t - T_1) ,
                       \label{dL1} \\
   \delta L_2(t) & = & \delta x_b(t - T_2) - \delta x_a(t) ,
                       \label{dL2}
\end{eqnarray}
where $T_1$ and $T_2$ are the photon propagation times for the forward
and return trip correspondingly. According to these definitions, the
displacement of one test mass is compared with the displacement of
the other at a later time to allow for finite delay from the light
propagation, as can be seen from Fig.~\ref{rtripLL}. Note that the
propagation times $T_1$ and $T_2$ in Eqs.(\ref{dL1})-(\ref{dL2}) can
be replaced with their nominal value $T$ because the test mass
displacements are already first order in $h$.

The total change in the distance between the masses in one round-trip 
of light would be 
\begin{eqnarray}
   \delta L_{\mathrm{r.t.}}(t) 
      & = & \delta L_1(t - T) + \delta L_2(t) \nonumber \\
      & = & 2 \, \delta x_b(t - T) - \delta x_a(t) - 
            \delta x_a(t - 2T) .
\end{eqnarray}
An explicit formula for this length change written in terms of the
amplitude of the gravitational wave is
\begin{equation}\label{dLcausal}
   \delta L_{\mathrm{r.t.}}(t) = (l + L) h(t - T) - 
      \frac{1}{2} \, l \, h(t) -  \frac{1}{2} \, l \, h(t - 2T) .
\end{equation}
This quantity represents causal length variations in laser
interferometers for gravitational wave detection. Note that $\delta
L_{\mathrm{r.t.}}$ is not translationally invariant, despite the fact
that it represents the relative displacement of the test masses. This
is the price one has to pay for satisfying the causality condition.

Changes in the distance, Eq.(\ref{dLcausal}), lead to changes in the 
round-trip time for photons propagating between the masses:
\begin{equation}\label{deltaTx}
   \frac{\delta_x T(t)}{T} = h(t - T) - 
      \mu \left[ h(t) - 2 h(t - T) + h(t - 2T) \right] ,
\end{equation}
where we introduced a dimensionless parameter 
\begin{equation}
    \mu = \frac{l}{2L} .
\end{equation}
The presence of this parameter in the subsequent formulas will
indicate the loss of translational invariance. The Laplace-domain 
version of Eq.(\ref{deltaTx}) can be written in a manner similar to 
Eq.(\ref{dTinTTh}), namely
\begin{equation}
   \frac{\delta_x \tilde{T}(s)}{T} = D_x(s) \, \tilde{h}(s) .
\end{equation}
where $D_x(s)$ is the corresponding response function 
\begin{equation}\label{respDx}
   D_x(s) = e^{-sT} - \mu \left( 1 - e^{-sT} \right)^2 .
\end{equation}
Note that $D_x(s)$ depends on the choice of the origin for this
coordinate system. At first it may seem that this loss of
translational invariance is natural. After all, the potential
explicitly depends on coordinates, which in classical mechanics
usually means that the symmetry with respect to translations is
lost. However, such a conclusion would  contradict our physical
intuition which maintains that all locations on the wavefront of the
plane gravitational wave must be equivalent. This implies that
physical quantities must be the same no matter where on this plane
they are measured, even though the potential explicitly discriminates
between different locations on the plane. We will see shortly that
this is indeed the case and that translational invariance is restored, 
but only when another significant effect is added to the picture: the 
gravitational redshift of light propagating between the masses.

\section{Distributed Gravitational Redshift}

We have calculated variations in the photon round-trip time which come
from the motion of the test masses induced by the gravitational wave.
In this calculation, we implicitly assumed that the propagation of
the photon between the test masses is uniform, as if it were moving
in flat space-time. However, the presence of the tidal forces
indicates that space-time is curved. As a result, the bouncing photon
will experience a gravitational redshift. There will be two such
effects in the following calculations. The first will be called the 
distributed gravitational redshift because it requires spacial
separation, the second will be called the localized gravitational 
redshift because it occurs at a single point in space.

The distributed gravitational redshift can be calculated as follows. 
Consider the interval for photons propagating along the $x$-axis:
\begin{equation}
   d s^2 = g_{00} \; c^2 \, dt^2 + dx^2 ,
\end{equation}
where $g_{00}$ is the time component of the metric, Eq.(\ref{g00postN}).   
The condition for a null trajectory ($ds = 0$) gives us the coordinate 
velocity of the photons:
\begin{equation}\label{photonVel}
   v^2 \equiv \left( \frac{dx}{dt} \right)^2 = c^2 + 2 \Phi(t,x) .
\end{equation}
To first order in $h$, the velocity can be approximated by
\begin{equation}
   v \approx \pm \; c \left[ 1 + \frac{1}{c^2} \Phi(t,x) \right],
\end{equation}
where $+$ and $-$ correspond to the forward and return trip, 
respectively.

Knowing the coordinate velocity of the photons, we can define the 
propagation time for the photon traveling between the masses:
\begin{equation}
   T_1(t) = \int\limits_{x_a(t-T_1)}^{x_b(t)} \frac{dx}{v} ,
            \qquad {\mathrm{and}} \qquad 
   T_2(t) = \int\limits_{x_b(t-T_2)}^{x_a(t)} \frac{(-dx)}{v} ,
\end{equation}
in accordance with Fig.~\ref{rtripLL}. We will not attempt to
calculate these integrals directly. Such calculations would be 
complicated because the boundaries of these integrals are changing
with time:
\begin{eqnarray}
   x_a(t) & = & l + \delta x_a(t) ,\\
   x_b(t) & = & l + L + \delta x_b(t).
\end{eqnarray}
Fortunately, we do not need to calculate the contributions of the 
boundary terms to these integrals. To first order in $h$, these 
contributions can be approximated by $\delta L_1(t)/c$ and 
$\delta L_2(t)/c$ (see Eqs.(\ref{dL1}) and (\ref{dL2})). Therefore, the 
combined effect of the varying boundaries is given by $\delta_x T$,
previously found in Eq.(\ref{deltaTx}). Thus, we only calculate the
times for photon propagation between the fixed boundaries: $l$ and $l+L$.
Such propagation times will be denoted here by $\Delta T_{1,2}$ to be 
distinguished from $T_{1,2}$.

In the forward trip, the propagation time between the fixed boundaries 
is
\begin{eqnarray}
   \Delta T_1(t) & = & \int\limits_l^{l + L} \frac{dx}{v(t',x)} \\
          & \approx & T - \frac{1}{c^3} \int\limits_l^{l + L}
                      \Phi(t', x) \, dx ,
\end{eqnarray}
where $t'$ is the retardation time which corresponds to the
unperturbed photon trajectory: $t' = t - (l + L - x)/c$. Similarly, 
the propagation time between the fixed boundaries in the return trip
is
\begin{equation}
   \Delta T_2(t) = T - \frac{1}{c^3} \int\limits_{l + L}^l 
                   \Phi(t', x) \, (-dx) ,
\end{equation}
though now the retardation time is given by $t' = t - (x - l)/c$. The 
round-trip time for photons traveling between the fixed boundaries can 
be found by adding $\Delta T_2(t)$ and $\Delta T_1(t-T)$. Deviations 
of this round-trip time from its unperturbed value, $2T$, are given by
\begin{equation}
   \delta_v T(t) = - \frac{1}{c^3} \int\limits_l^{l + L} \left[
      \Phi \left( t - 2T + \frac{x-l}{c}, x \right) + 
      \Phi \left( t - \frac{x-l}{c}, x \right) \right] dx .
      \label{dTinFNPhi}
\end{equation}
After replacing the potential with its explicit form, 
Eq.(\ref{potential}), we obtain the formula for $\delta_v T$ in terms 
of the amplitude of the gravitational wave:
\begin{equation}\label{deltaTv}
   \delta_v T(t) = \frac{1}{4c^3} \int\limits_l^{l + L} \left[
      \ddot{h}\left( t - 2T + \frac{x-l}{c} \right) + 
      \ddot{h}\left( t - \frac{x-l}{c} \right) \right] x^2 \, dx .
      \label{dTinFNh}
\end{equation}
This quantity represents the effect of the distributed gravitational 
redshift \cite{Rakhmanov:PhD}.

Equation (\ref{dTinFNh}) bears close similarity with
Eq.(\ref{delayTT}), as both formulas represent cumulative effects 
of the gravitational wave. However, unlike Eq.(\ref{delayTT}), which
is translationally invariant, Eq.(\ref{dTinFNh}) is not, as can be seen
from the presence of $x^2$-factor in the integrand. A better way to 
analyze the loss of translational invariance would be to rewrite the 
result in the Laplace domain:
\begin{equation}
   \frac{\delta_v \tilde{T}(s)}{T} = D_v(s) \, \tilde{h}(s) ,
\end{equation}
where $D_v(s)$ is the corresponding response function:
\begin{eqnarray}
   D_v(s) & = & \frac{1}{2sT} \left( 1 - e^{-2sT} \right) - 
                e^{-sT} + \nonumber \\
          &   & \mu \left( 1 - e^{-sT} \right)^2 + 
                \mu^2 \left( 1 - e^{-2sT} \right) s T .
                \label{respDt}
\end{eqnarray}
The terms proportional to $\mu$ and $\mu^2$ represent the dependence
of the response function on the choice of the origin for this
coordinate system.

We can now combine the variations in the photon propagation time which
are caused by the motion of the test masses with those caused by the 
distributed gravitational redshift. The resulting round-trip time
would be
\begin{equation}\label{TrtXV}
   T_{\mathrm{r.t.}} = 2T + \delta_x T + \delta_v T .
\end{equation}
To this point, the combined effect of the gravitational wave is 
given by the sum:
\begin{equation}\label{DxDv}
   D_x(s) + D_v(s) = \left( \frac{1}{2sT} + \mu^2 s T \right)
      \left( 1 - e^{-2sT} \right).
\end{equation}
By adding the two response functions we cancel the terms proportional 
to $\mu$. However, the term proportional to $\mu^2$ remains. As will
be shown next, this term is related to the localized gravitational
redshift.

\section{Localized Gravitational Redshift}

The last contribution to the photon round-trip time is also related to
the gravitational redshift, although it is somewhat different from the 
distributed effect described above. In the presence of gravitational 
waves the clocks at different places run differently. The rate
($dt^*$) of the clock which is located at $x$ is related to the rate 
($dt$) of the clock at the origin by 
\begin{equation}
   {d t^*}^2 = - g_{00}(t,x) \; dt^2 ,
\end{equation}
which is the proper time at this location. In the above calculation
of the photon round-trip time, Eq.(\ref{TrtXV}), we implicitly
assumed that the time is measured with the clock at the origin:
$x=0$. The photon trajectories, however, begin and end at the location
of the first test mass, a finite distance ($l$) away from the origin.
As a result, the readings of time become dependent on this distance. 
To avoid this problem, we shall measure time with the clock located 
at $x=l$. For this clock, the round-trip time is different from 
$T_{\mathrm{r.t.}}$, Eq.(\ref{TrtXV}). The presence of the
time-dependent gravitational potential affects the reading of this 
clock, causing it to register the round-trip time as
\begin{eqnarray}
   T_{\mathrm{r.t.}}^*(t) 
      & = & \int\limits_{t - T_{\mathrm{r.t.}}}^{t}
            \sqrt{-g_{00}(t',l)} \, dt' \nonumber \\
      & \approx & T_{\mathrm{r.t.}}(t) + \frac{1}{c^2} 
                  \int\limits_{t - T_{\mathrm{r.t.}}}^{t}
                  \Phi(t',l) \, dt'. \label{localizedGR} 
\end{eqnarray}
To first order in $h$, the variation of the round-trip time due to
this effect can be estimated as
\begin{eqnarray}
   \delta_t T(t)
      & \approx & \frac{1}{c^2} \int\limits_{t-2T}^{t} 
            \Phi(t',l) \, dt'  \label{deltaTt} \\
      & = & - \frac{l^2}{4c^2} \left[ \dot{h}(t) - 
            \dot{h}(t - 2T) \right] .
\end{eqnarray}
This contribution to the round-trip propagation time comes from the 
non-uniformity of time flow caused by the presence of the gravitational 
wave. It will be called here the localized gravitational redshift. In 
the Laplace domain it can be written as
\begin{equation}
   \frac{\delta_t \tilde{T}(s)}{T} = D_t(s) \, \tilde{h}(s) ,
\end{equation}
where $D_t(s)$ is the corresponding response function
\begin{equation}
   D_t(s) = - \mu^2 \left( 1 - e^{-2sT} \right) s T .
\end{equation}
Addition of this response function to Eq.(\ref{DxDv}) will cancel 
the $\mu^2$-terms, giving us a translationally invariant result.

We can now conclude that the change in the round-trip time caused 
by the gravitational wave consists of three contributions:
\begin{equation}\label{dTsum}
   \delta T = \delta_x T + \delta_v T + \delta_t T ,
\end{equation}
which come from displacement of the test masses, changes in the
coordinate velocity of the photons and variations in the clock rate. 
The combined result of these effects is given by the sum:
\begin{equation}\label{Hcomb}
   D_x(s) + D_v(s) + D_t(s) = \frac{1 - e^{-2sT}}{2sT} ,
\end{equation}
which is translationally invariant. Furthermore, the sum gives us a
response function which is identical to $C(s)$, Eq.~(\ref{respTT}),
which proves that the two coordinate systems indeed yield the same 
answer for the observable photon round-trip time.

The requirement of translational invariance played a special role in
the above analysis. The coordinate transformations,
Eq.(\ref{shiftsXY}), are a particular case of transformations known 
as changes of the origin, which in general relativity are 
accomplished with the help of parallel transports \cite{Synge:1960}.
Following the Newtonian style of our presentation, we referred  
to these transformations as {\emph{translations}} and assumed that
they represent a symmetry. The origin of this symmetry 
is related to the planeness of the gravitational wave
\cite{Bondi:1959}.

\section{The Round-Trip Phase of Light}

In the above picture, we considered the bouncing photon as a particle,
assuming that there is a beginning and an end for the photon round
trips. In practice, measurements of photon propagation times are
usually done with optical interferometry in which photons are
represented by continuous electromagnetic waves. We shall therefore
briefly describe how the above calculations can be modified to become
applicable to continuous waves. To be specific, we assume that the
light is represented by a plane monochromatic wave with frequency
$\omega$ and wavenumber $k$. In the absence of gravitational waves,
such a wave is given by $\exp[i (\omega t \mp k x) ]$. Then the
photon trajectory introduced above would describe advancement of
a surface of constant phase, whereas the photon velocity becomes 
the phase velocity of the wave. In this approach, the quantity of
interest would be the round-trip phase, or more precisely, its
variation caused by the gravitational wave.

The first contribution to the round-trip phase comes from the motion
of the test masses:
\begin{equation}\label{DeltaPsi}
   \psi_x = - k \, \delta L_{\mathrm{r.t.}} = 
            - \omega \; \delta_x T ,
\end{equation}
where $\delta L_{\mathrm{r.t.}}$ represents variations in the distance
between the test masses, Eq.(\ref{dLcausal}), and $\delta_x T $ 
represents the corresponding time variations, Eq.(\ref{deltaTx}).
The second contribution comes from the variations in the phase
velocity of the wave:
\begin{equation}\label{psi_k}
   \psi_k = \frac{k}{c^2} \int\limits_{\mathcal{C}} \Phi \, dx . 
\end{equation}
Here we give a brief derivation of this result based on simple
physical arguments. (Another derivation, based on the solution of the
eikonal equation, is given  in Appendix \ref{App:eikonal}.)

In the presence of gravitational waves, the frequency and wavenumber
are no longer constant; they become functions of position and time:
$\Omega(x,t)$ and $K(x,t)$. Then the dispersion relation for the 
electromagnetic wave would read
\begin{equation}\label{dispersion}
   \Omega^2 = v^2 \, K^2 ,
\end{equation}
where $v$ is the phase velocity of the light, defined in 
Eq.(\ref{photonVel}). To first order in $h$, Eq.(\ref{dispersion}) 
can be written as
\begin{equation}\label{linDisp}
   \Omega - c \, K = \frac{k}{c} \; \Phi .
\end{equation}
For a plane electromagnetic wave moving in the positive $x$-direction, 
an infinitesimal phase change is given by $(\Omega \, dt - K \, dx)$. 
Then the accumulated phase change can be found by integrating this 
quantity along the trajectory of a given wavefront. In doing so, we 
would find that the accumulated phase change vanishes because 
$dx/dt = \Omega/K$. This result is quite natural, as traveling with
the wavefront means following the surface of constant phase for which 
no phase change ensues. However, we must remember that we are not 
interested in the absolute phase change along the photon
trajectory. Rather, we are interested in the variation of this phase 
change with respect to the unperturbed wave. Such a phase variation
would be given by 
\begin{equation}
   \psi_k = \int\limits_{\mathcal{C}} (\Omega \, dt - K \, dx),
\end{equation}
where $\mathcal{C}$ denotes the unperturbed trajectory: $dx/dt=\pm c$. 
Taken along the unperturbed photon trajectory, such an integral would
yield a non-zero answer, which is equivalent to Eq.(\ref{psi_k}). 
Note that the integral over the photon trajectory, Eq.(\ref{psi_k}), 
has already been calculated, see Eq.(\ref{dTinFNPhi}). Therefore,
the phase change can then be written as
\begin{equation}
   \psi_k = - \omega \; \delta_v T ,
\end{equation}
where $\delta_v T$ is the corresponding variation in the round-trip
time.

We now can add this phase change to the phase change produced by the 
motion of the test masses, Eq.(\ref{DeltaPsi}). There is no need to 
worry about the difference between $k$ and $K$ for this part. The 
displacements of the test masses are first order in $h$, and therefore 
any correction to $k$ would result in second order terms. Thus, the
combined effect is given by
\begin{equation}\label{psiXpsiK}
   \psi = \psi_x + \psi_k = - \omega \, (\delta_x T + \delta_v T).
\end{equation}
As we already know from Eqs.(\ref{TrtXV}) and (\ref{DxDv}), this
quantity is not translationally invariant, which means that it cannot
be observed in the experiment. The change in the phase shift $\psi$ 
represents the difference of the round-trip phases for two different
electromagnetic waves: with and without the gravitational wave. 
Such a phase change cannot be measured in the experiment 
as the two waves cannot exist in the same space-time. To form an
observable quantity, we shall compare the phase change of the
probe electromagnetic wave with that of a reference wave. The 
natural reference is the source itself, and therefore we need to find
the phase change of the  source.

In flat space-time, the phase of the source would simply be 
$\omega t$, and the phase shift of the source $2 \omega T$. In the 
presence of gravitational waves, the phase of the source becomes
$\omega t^*$, where $t^*$ is the proper time at the location of the
source. Then the phase shift of the source can be found as
\begin{equation}
   \omega \left[ t^*(t) - t^*(t - T_{\mathrm{r.t.}}) \right] = 
      \omega \, T_{\mathrm{r.t.}}^*(t) .
\end{equation}
Therefore, the change in this phase shift which is produced by the 
gravitational wave is 
\begin{equation}\label{psi_source}
   \psi_{\mathrm{so}} = \omega \; \delta_t T ,
\end{equation}
where $\delta_t T$ is given by Eq.(\ref{deltaTt}). 

We can now compare the phase change of the moving wave, 
Eq.(\ref{psiXpsiK}), with that of a static source, 
Eq.(\ref{psi_source}). The difference between the phase change of the
wavefront for the electromagnetic wave returning to the source and the
phase change of the source at that moment is
\begin{equation}
   \delta \psi = \psi - \psi_{\mathrm{so}} .
\end{equation}
In the explicit form this phase difference is given by
\begin{equation}
   \delta \psi = - \omega (\delta_x T + \delta_v T + \delta_t T) .
\end{equation}
As we already know from Eqs.(\ref{dTsum}) and (\ref{Hcomb}), this
phase is translationally invariant and therefore represents an
observable quantity. It is not surprising that this phase deviation 
is related to the time deviation by a simple formula:
\begin{equation}
   \delta \psi = - \omega \; \delta T .
\end{equation}
This result could have been guessed from simple dimensional analysis
and the requirement of translational invariance. The above derivation 
serves to explain the physical meaning of the relative phase shift and 
its constituent parts. In short, the motion of the test masses and the 
distributed gravitational redshift appear now as the phase shift of 
the traveling wave, whereas the localized gravitational redshift
appears as the phase shift of the static source.

\section{Concluding Remarks}

We have shown that the response of test masses to gravitational waves
in the local Lorentz gauge acquires contributions from three different
effects: the motion of the test masses and the distributed and
localized gravitational redshifts. Only when taken together do these
effects yield an observable quantity. The approach developed in this
paper has  allowed us to calculate physical quantities directly in the
coordinates of the local observer. In these coordinates, the
assumption of the test mass inertiality is not required, and various
forces acting on the masses can be added at will. We have provided a
consistent framework for doing calculations in the coordinate system
which is more natural for ground-based laser gravitational-wave
detectors than the TT-gauge.

To simplify the calculations, we introduced the three effects of 
the gravitational wave in a step-by-step fashion. At each step,
mathematical derivations took advantage of the previous step. In
retrospect, it is clear that a more direct way of doing the
calculations would be to start with an abstract definition:
\begin{equation}
   T_{\mathrm{r.t.}}^* = \int dt^* ,
\end{equation}
and then to proceed with integration over the photon trajectory
\begin{equation}
   T_{\mathrm{r.t.}}^*
   = \int\limits \sqrt{-g_{00}(l, t')} \, dt'
   = \int\limits_{\mathcal{C}^*} \sqrt{-g_{00}(l, t')} \, 
     \frac{dx}{v(x, t')} .
\end{equation}
In this approach the contour $\mathcal{C}^*$ would represent the actual 
photon trajectory: $dx/dt = \pm v(x,t)$ which extends to the actual
test mass positions: $l + \delta x_a$ and $l + L + \delta x_b$. 
By evaluating various terms in the contour integral to first order in
$h$, one would reproduce the above three contributions to the
round-trip time. Although this approach may seem different from the 
one described in this paper, the mathematical equations and their
physical interpretations would be essentially the same.

\section*{Acknowledgments}

I am indebted to Barry Barish who encouraged this research as a part
of my thesis project. Also I would like to thank Kip Thorne and
Bernard Whiting for fruitful discussions and for comments on the draft
of this paper. This research was supported in part by the National
Science Foundation under grants PHY-9210038 and PHY-0070854.

\appendix   
\section{Coordinate and Metric Transformations}
\label{App:TT2FN}

For completeness, we present here the transformations from the
TT-gauge to the gauge of a local observer. Denote the coordinates of a 
local observer by $x^{\mu}$ and the metric in these coordinates by 
$g_{\mu \nu}$. Also, denote the coordinates of the TT-gauge by 
$\bar{x}^{\mu}$ and the corresponding metric by 
$\bar{g}_{\alpha \beta}$. The components of the metric in the 
TT-gauge, Eq.(\ref{metricTT}), can be written as
\begin{equation}\label{weakGW}
   \bar{g}_{\mu \nu} = \eta_{\mu \nu} + \left(
   \begin{array}{cccc}
   0  &  0  &  0  &  0  \\
   0  &  h  &  0  &  0  \\
   0  &  0  & -h  &  0  \\
   0  &  0  &  0  &  0 
   \end{array}\right) ,
\end{equation}
where $\eta_{\mu \nu} = \mathrm{diag}\{-1,1,1,1 \}$ is the Minkowski
metric, and $h = h(\bar{t} + \bar{z}/c)$. From general relativity we
know that the coordinate transformations,
$\bar{x}^{\mu} \rightarrow x^{\mu}$, induce the transformations of 
the metric:
\begin{equation}\label{eq:g2g}
   g_{\mu \nu} = 
      \frac{\partial \bar{x}^{\alpha}}{\partial x^{\mu}}  \;
      \frac{\partial \bar{x}^{\beta}} {\partial x^{\nu}}  \;\, 
      \bar{g}_{\alpha \beta} .
\end{equation}
By definition, $g_{\mu \nu}$ must become the Minkowski metric at the 
origin, and all its derivatives must vanish at this point. There are 
a number of metrics which satisfy these conditions. Here we consider
one such choice \cite{Grishchuk:1977, Grishchuk:1980, Pegoraro:1978}. 
It can be obtained with the coordinate transformations, which to first
order in $h$, are given by
\begin{eqnarray}
   \bar{t} & = & t - \frac{1}{4c^2} \, \dot{h} \,
           (x^2 - y^2) ,
           \label{t2t} \\
   \bar{x} & = & x - \frac{1}{2} \, h \, x ,
           \label{x2x} \\
   \bar{y} & = & y + \frac{1}{2} \, h \, y ,
           \label{y2y} \\
   \bar{z} & = & z + \frac{1}{4c}  \, \dot{h} \,
           (x^2 - y^2).
           \label{z2z}
\end{eqnarray}
The corresponding metric tensor can be obtained by performing the 
induced transformation, Eq.(\ref{eq:g2g}). To first order in $h$, 
the result is
\begin{equation}\label{exactSol}
   g_{\mu \nu} = \eta_{\mu \nu} - \frac{2}{c^2} \left(
   \begin{array}{cccc}
   \Phi  &  0  &  0  & \Phi  \\
      0  &  0  &  0  &  0    \\
      0  &  0  &  0  &  0    \\
   \Phi  &  0  &  0  & \Phi
   \end{array}\right) ,
\end{equation}
where $\Phi$ is a function of the new coordinates:
\begin{equation}
   \Phi = - \frac{1}{4} \, \ddot{h}(t + z/c) \, (x^2 - y^2) .
\end{equation}
In the post-Newtonian approach, this function becomes the potential 
for the tidal forces, which for $z=0$ appeared in Eq.(\ref{potential}).

It is interesting to note that although the transformation rules
Eqs.(\ref{t2t})-(\ref{z2z}) are approximate, the metric 
Eq.(\ref{exactSol}) is an exact solution of Einstein equations 
\cite{Thorne:2003}. This metric is generally known as the
plane-front solution for strong gravitational waves 
\cite{Peres:1959, Ehlers:1962}. Further discussion of the relationship 
between the metric of the local observer and the exact solution 
can be found in \cite{Blandford:2003}.

The propagation of an electromagnetic wave in curved space-time is 
described by the eikonal $\Psi$ \cite{Landau:1971}, which satisfies 
the equation:
\begin{equation}
   g^{\mu \nu} \frac{\partial \Psi}{\partial x^{\mu}} 
               \frac{\partial \Psi}{\partial x^{\nu}} = 0 ,
\end{equation}
where $g^{\mu \nu}$ stands for the contravariant metric tensor. 
Its components are given by
\begin{equation}
   g^{\mu \nu} = \eta^{\mu \nu} - \frac{2}{c^2} \left(
   \begin{array}{cccc}
   -\Phi &  0  &  0  & \Phi  \\
      0  &  0  &  0  &  0    \\
      0  &  0  &  0  &  0    \\
   \Phi  &  0  &  0  & -\Phi
   \end{array}\right) .
\end{equation}
For light propagating along the $x$-axis, the eikonal equation becomes
\begin{equation}\label{eikonalEq}
   \left( 1 - \frac{2}{c^2} \, \Phi \right)
      \left( \frac{\partial \Psi}{\partial ct} \right)^2 =
      \left( \frac{\partial \Psi}{\partial x} \right)^2 .
\end{equation}
The solution of this equation is described next.

\section{Solution to the Eikonal Equation}
\label{App:eikonal}

\begin{figure}[t]
   \centering\includegraphics[width=0.6\textwidth]{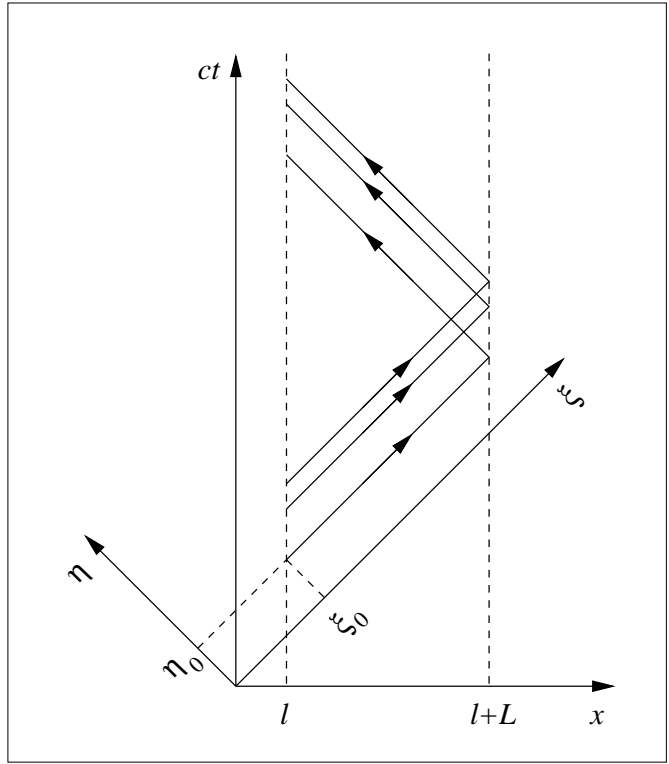}
   \caption{Propagation of the electromagnetic waves in the light 
   cone coordinates. The source is located at $x=l$ and the turning 
   point at $x=l+L$.}
   \label{lightcone}
\end{figure}

The eikonal equation can reduced to a linear differential equation 
by taking the square root of both sides of Eq.(\ref{eikonalEq}) and 
by keeping only the terms which are first order in $h$:
\begin{equation}\label{eikonalEqLin}
    \left( \frac{\partial}{\partial ct} \pm \, 
           \frac{\partial}{\partial x} \right) \Psi = 
           \frac{1}{c^2} \; \Phi \; \left(
           \frac{\partial \Psi}{\partial ct} \right),
\end{equation}
where $\pm$ correspond to the wave propagation in the positive and 
negative $x$-directions. The large unperturbed value of the eikonal 
satisfies Eq.~(\ref{eikonalEqLin}) in the absence of the 
gravitational waves ($\Phi=0$), and is given by $\omega t \mp k x$
up to an additive constant. Therefore, to first order in $h$, the 
solution of the eikonal equation can be found as
\begin{eqnarray}
   \Psi_1 & = & \omega t - k x + k l + \delta \Psi_1 ,
                \label{DefPsi1} \\
   \Psi_2 & = & \omega t + k x - k(l + 2L) + \delta \Psi_2 ,
                \label{DefPsi2}
\end{eqnarray}
where $\delta \Psi_{1,2}$ are the first order perturbations. 
For convenience we introduce the light-cone coordinates:
\begin{eqnarray}
   \xi  & = & (ct + x)/2 , \\
   \eta & = & (ct - x)/2 ,
\end{eqnarray}
in which the photon world-lines become collinear with the $\xi$ and 
$\eta$ axes, as shown in Fig.~\ref{lightcone}. In these coordinates, the 
first order perturbations satisfy the equations:
\begin{eqnarray}
   \frac{\partial}{\partial \xi} \; \delta \Psi_1 & = & 
      \frac{k}{c^2} \; \Phi ,\\
   \frac{\partial}{\partial \eta} \; \delta \Psi_2 & = & 
      \frac{k}{c^2} \; \Phi .
\end{eqnarray}
These equations allow direct integration:
\begin{eqnarray}
   \delta \Psi_1(\xi, \eta) & = & 
      \frac{k}{c^2} \int\limits_{\xi_0}^{\xi} 
      \Phi(\xi', \eta) \, d\xi' + f_1(\eta) ,\\
   \delta \Psi_2(\xi, \eta) & = & 
      \frac{k}{c^2} \int\limits_{\eta_0}^{\eta} 
      \Phi(\xi, \eta') \, d\eta' + f_2(\xi) ,
\end{eqnarray}
where $f_1(\eta)$ and $f_2(\xi)$ are arbitrary at this point. 
Transforming back to the coordinates $x$ and $t$, we obtain the
solution:
\begin{eqnarray}
   \delta \Psi_1(x,t) & = & 
      \frac{k}{c^2} \int\limits_l^x \Phi \left(x', t - 
      \frac{x-x'}{c}\right) \, dx' + f_1(x,t) ,\\
   \delta \Psi_2(x,t) & = & 
      \frac{k}{c^2} \int\limits_x^{l+L} \Phi \left(x', t + 
      \frac{x-x'}{c}\right) \, dx' + f_2(x,t) .
\end{eqnarray}

The function $f_1$ is defined by the value of the eikonal at the
source for the electromagnetic wave: $\Psi_1(l,t) = \omega \, t^*$.
To first order in $h$, this value can be found as
\begin{eqnarray}
   \Psi_1(l,t) & = & \omega \int\limits_0^t 
      \sqrt{-g_{00}(l, t')} \; dt' \\
     & \approx & \omega \int\limits_0^t \left[ 1 + 
      \frac{1}{c^2} \Phi(l, t') \right] dt' ,
\end{eqnarray}
which must be the same as $\omega t + f_1(l, t)$, according to the
definition Eq.(\ref{DefPsi1}). We thus find the function $f_1$ at the 
location of the source. Knowing that $f_1$ is a function of $ct-x$, we 
can extend the values of $f_1$ from the source location to the entire 
$xt$-plane:
\begin{equation}
   f_1(x,t) = \frac{k}{c} \int\limits_0^{t - \frac{x-l}{c}} 
      \Phi(l, t') \; dt' .
\end{equation}
The function $f_2$ is defined by the value of the eikonal $\Psi_1$ 
at the turning point. The continuity of the eikonal implies that 
\begin{equation}
   \delta \Psi_1(l + L,t) = \delta \Psi_2(l + L,t) .
\end{equation}
From this condition we can find $f_2$ at the turning point. Knowing 
that $f_2$ is a function of $ct+x$, we can extend the values of
$f_2$ from the turning point to the entire $xt$-plane:
\begin{eqnarray}
   f_2(x,t) & = & \frac{k}{c} \int\limits_0^{t - 2T + 
                  \frac{x-l}{c}} \Phi(l, t') \; dt' + \nonumber \\
            &   & \frac{k}{c^2} \int\limits_l^{l + L} \Phi \left(
                  x', t - 2T + \frac{x + x' - 2l}{c}\right) \, dx' .
\end{eqnarray}

The phase shift acquired by the electromagnetic wave in one round
trip is given by the difference between the value of the eikonal at
the beginning and the end of the propagation. To first order in $h$,
this phase shift is given by
\begin{equation}
   \psi_k(t) = \delta \Psi_2(l, t) - \delta \Psi_1(l, t - 2T) .
\end{equation}
Simple algebra shows that this definition leads to
\begin{eqnarray}
   \psi_k(t) & = & \frac{k}{c^2} \int\limits_l^{l+L} \Phi 
       \left(x, t - \frac{x - l}{c} \right) \, dx + \nonumber \\
             &   & \frac{k}{c^2} \int\limits_l^{l+L} \Phi 
       \left(x, t - 2T + \frac{x - l}{c} \right) \, dx ,
\end{eqnarray}
which is an extended form of Eq.(\ref{psi_k}).

Finally, we give explicit formulas for $\Omega$ and $K$ in
terms of the gravitational potential. These two quantities can be
derived from the eikonal:
\begin{equation}
   \Omega = \frac{\partial \Psi}{\partial t}, 
      \qquad {\mathrm{and}} \qquad 
   K = \mp \frac{\partial \Psi}{\partial x} ,
\end{equation}
where $-$ corresponds to the forward trip and $+$ to the return
trip. For example, in the forward propagation
\begin{eqnarray}
   \Omega(x,t) & = & \omega + \frac{k}{c} \, \Gamma(x,t) + 
      \frac{k}{c} \, \Phi\left(l,t - \frac{x - l}{c} \right) ,\\
   K(x,t) & = &  k + \frac{k}{c^2} \, \Gamma(x,t) + 
      \frac{k}{c^2} \, \Phi\left(l,t - \frac{x - l}{c} \right) 
       - \frac{k}{c^2} \, \Phi(x,t) ,
\end{eqnarray}
where $\Gamma$ represents the non-stationary effect of the
gravitational redshift:
\begin{equation}
   \Gamma(x,t) = \frac{1}{c} \int\limits_l^x 
      \frac{\partial}{\partial t} \left[ \Phi \left(x', t - 
      \frac{x - x'}{c} \right) \right] \, dx' .
\end{equation}
Note that $K$ can also be written as
\begin{equation}
   K(x,t) = \frac{1}{c} \, \Omega(x,t) - \frac{k}{c^2} \, \Phi(x,t) ,
\end{equation}
which leads directly to the dispersion relation, Eq.(\ref{linDisp}).


\begin{thebibliography}{10}

\bibitem{Barish:1999}
B.~Barish and R.~Weiss, ``{LIGO} and the detection of gravitational waves,''
  {\em Physics Today}, vol.~52, pp.~44--50, 1999.

\bibitem{Bradaschia:1990}
C.~Bradaschia {\em et~al.}, ``The {VIRGO Project} -- a wide band antenna for
  gravitational-wave detection,'' {\em Nuclear Instruments and Methods in
  Physics Research A}, vol.~289, pp.~518--525, 1990.

\bibitem{Misner:1973}
C.~Misner, K.~Thorne, and J.~Wheeler, {\em Gravitation}.
\newblock San Francisco: W.H. Freeman and Company, 1973.

\bibitem{Buonanno:2001}
A.~Buonanno and Y.~Chen, ``Quantum noise in second generation, signal-recycled
  laser interferometric gravitational-wave detectors,'' {\em Physical Review
  D}, vol.~64, 2001.
\newblock 042006.

\bibitem{Khalili:2001}
F.~Khalili, ``Frequency-dependent rigidity in large-scale interferometric
  gravitational-wave detectors,'' {\em Physics Letters A}, vol.~288,
  pp.~251--256, 2001.

\bibitem{Rakhmanov:PhD}
M.~Rakhmanov, {\em Dynamics of Laser Interferometric Gravitational Wave
  Detectors}.
\newblock PhD thesis, California Institute of Technology, 2000.

\bibitem{Weber:1961}
J.~Weber, {\em General Relativity and Gravitational Waves}.
\newblock New York: Interscience Publishers, Inc., 1961.

\bibitem{Grishchuk:1977}
L.~Grishchuk, ``Gravitational waves in the cosmos and the laboratory,'' {\em
  Soviet Physics, Uspekhi}, vol.~20, pp.~319--334, 1977.

\bibitem{Pegoraro:1978}
F.~Pegoraro, E.~Picasso, and L.~Radicati, ``On the operation of a tunable
  electromagnetic detector for gravitational waves,'' {\em Journal of Physics
  A: Mathematical and General}, vol.~11, pp.~1949--1962, 1978.

\bibitem{Fortini:1982}
P.~Fortini and C.~Gualdi, ``Fermi normal coordinate system and electromagnetic
  detectors of gravitational waves,'' {\em Il Nuovo Cimento}, vol.~71 B,
  pp.~37--54, 1982.

\bibitem{Flores:1986}
G.~Flores and M.~Orlandini, ``Comparison between the fermi normal and the
  transverse traceless co-ordinate system,'' {\em Il Nuovo Cimento}, vol.~91 B,
  pp.~236--240, 1986.

\bibitem{Grishchuk:1980}
L.~Grishchuk and A.~Polnarev, ``Gravitational waves and their interaction with
  matter and fields,'' in {\em General Relativity and Gravitation: One hundred
  years after the birth of Albert Einstein} (A.~Held, ed.), vol.~2,
  pp.~393--434, New York: Plenum Press, 1980.

\bibitem{Fortini:1990}
P.~Fortini and A.~Ortolan, ``Some remarks on electromagnetic detectors of
  gravitational waves,'' in {\em Problems of fundamental modern physics:
  proceedings of the 4th Winter School on Hadronic Physics} (B.~M.
  Roberto~Cherubini, Pietro~Dalpiaz, ed.), (Folgaria (Trento), Italy),
  pp.~468--478, World Scientific, 1990.

\bibitem{Callegari:1987}
G.~Callegari, P.~Fortini, and C.~Gualdi, ``On the crucial role played by the
  reference system in gravitational-wave detector theory,'' {\em Il Nuovo
  Cimento}, vol.~100 B, pp.~421--424, 1987.

\bibitem{Fortini:1991}
P.~Fortini and A.~Ortolan, ``Light phase shift in the field of a gravitational
  wave,'' {\em Il Nuovo Cimento, Note Brevi}, vol.~106 B, pp.~101--104, 1991.

\bibitem{Schutz:1985}
B.~Schutz, {\em A first course in general relativity}.
\newblock Cambridge: Cambridge University Press, 1985.

\bibitem{Synge:1960}
J.~Synge, {\em Relativity: The General Theory}.
\newblock Amsterdam: North-Holland Publishing Company, 1960.

\bibitem{Mizuno:1997}
J.~Mizuno, A.~R\"{u}diger, R.~Schilling, W.~Winkler, and K.~Danzmann,
  ``Frequency response of {M}ichelson- and {S}agnac-based interferometers,''
  {\em Optics Communications}, vol.~138, pp.~383--393, 1997.

\bibitem{Estabrook:1975}
F.~B. Estabrook and H.~D. Wahlquist, ``Response of {D}oppler spacecraft
  tracking to gravitational radiation,'' {\em General Relativity and
  Gravitation}, vol.~6, pp.~439--447, 1975.

\bibitem{Estabrook:1985}
F.~Estabrook, ``Response functions of free mass gravitational wave antennas,''
  {\em General Relativity and Gravitation}, vol.~17, pp.~719--724, 1985.

\bibitem{Gursel:1984}
Y.~G\"{u}rsel, P.~Linsay, R.~Spero, P.~Saulson, S.~Whitcomb, and R.~Weiss,
  ``Response of a free mass interferometric antenna to gravitational wave
  excitation,'' in {\em A Study of a Long Baseline Gravitational Wave Antenna
  System}, National Science Foundation Report, 1984.

\bibitem{Vinet:1986}
J.-Y. Vinet, ``Recycling interferometric antennas for periodic gravitational
  waves,'' {\em Journal De Physique}, vol.~47, pp.~639--643, 1986.

\bibitem{Saulson:1994}
P.~Saulson, {\em Fundamentals of Interferometric Gravitational Wave Detectors}.
\newblock Singapore: World Scientific, 1994.

\bibitem{Blandford:2003}
R.~Blandford and K.~S. Thorne, {\em Ph 136: Applications of Classical Physics},
  ch.~26.
\newblock California Institute of Technology, Pasadena, 2003.
\newblock available on line at
  ``http://www.pma.caltech.edu/Courses/ph136/yr2002".

\bibitem{Bondi:1959}
H.~Bondi, F.~Pirani, and I.~Robinson, ``Gravitational waves in general
  relativity {III}. {E}xact plane waves,'' {\em Proceedings of the Royal
  Society of London. A}, vol.~251, pp.~519--533, 1959.

\bibitem{Markowicz:2003}
J.~Markowicz, R.~Savage, and P.~Schwinberg, ``Development of a readout scheme
  for high-frequency gravitational waves,'' {LIGO} technical document,
  California Institute of Technology, Pasadena, California, 2003.

\bibitem{Thorne:2003}
K.~S. Thorne, October 2003.
\newblock private communication.

\bibitem{Peres:1959}
A.~Peres, ``Some gravitational waves,'' {\em Physical Review Letters}, vol.~3,
  pp.~571--572, 1959.

\bibitem{Ehlers:1962}
J.~Ehlers and W.~Kundt, ``Exact solutions of the gravitational field
  equations,'' in {\em Gravitation: an introduction to current research}
  (L.~Witten, ed.), pp.~49--101, New York, London: John Wiley \& Sons, Inc.,
  1962.

\bibitem{Landau:1971}
L.~Landau and E.~Lifshitz, {\em The Classical Theory of Fields}.
\newblock New York: Pergamon Press, 1971.

\end{thebibliography}
\end{document}